\newlength{\absize}
\documentclass[12pt]{article}
\renewcommand{\baselinestretch}{1.5}
\renewcommand{\arraystretch}{1.5}
\begin{document}
\thispagestyle{empty}
\pagestyle{empty}
\renewcommand{\thefootnote}{\fnsymbol{footnote}}
\newcommand{\starttext}{\newpage\normalsize
\pagestyle{plain}
\setlength{\baselineskip}{3ex}\par
\setcounter{footnote}{0}
\renewcommand{\thefootnote}{\arabic{footnote}}
}
\newcommand{\preprint}[1]{\begin{flushright}
\setlength{\baselineskip}{3ex}#1\end{flushright}}
\renewcommand{\title}[1]{\begin{center}\LARGE
#1\end{center}\par}
\renewcommand{\author}[1]{\vspace{2ex}{\Large\begin{center}
\setlength{\baselineskip}{3ex}#1\par\end{center}}}
\renewcommand{\thanks}[1]{\footnote{#1}}
\renewcommand{\abstract}[1]{\vspace{2ex}\normalsize\begin{center}
\centerline{\bf Abstract}\par\vspace{2ex}\parbox{\absize}{#1
\setlength{\baselineskip}{2.5ex}\par}
\end{center}}

\newcommand{\rep}{representation}
\newcommand{\tr}{\mathop{\rm tr}}
\newcommand{\cO}{{\cal O}}
\newcommand{\cL}{{\cal L}}
\newcommand{\half}{{1\over2}}
\newcommand{\gtrsim}
{\raisebox{.2em}{$\rlap{\raisebox{-.5em}{$\;\sim$}}>\,$}}
\newcommand{\ltsim}
{\raisebox{.2em}{$\rlap{\raisebox{-.5em}{$\;\sim$}}<\,$}}
\newlength{\eqnparsize}
\setlength{\eqnparsize}{.95\textwidth}
\newcommand{\eqnbox}[1]{\parbox{\eqnparsize}{\bf\vskip.25ex
#1\vskip1ex}}
\newcommand{\PSbox}[3]{\mbox{\rule{0in}{#3}\includegraphics{#1}\hspace{#2
}}}
\def\spur#1{\mathord{\not\mathrel{#1}}}
\newcommand\etal{{\it et al.}}
\def\arctanh{\mathop{\rm arctanh}\nolimits}
\def\sech{\mathop{\rm sech}\nolimits}
\newcommand{\Ket}[1]{\left| #1 \right\rangle}

\setlength{\jot}{1.5ex}
\newcommand{\figsize}{\small}
\renewcommand{\bar}{\overline}
\font\fiverm=cmr5
\input prepictex
\input pictex
\input postpictex
\input{psfig.sty}
\newdimen\tdim
\tdim=\unitlength
\def\stpltsmbl{\setplotsymbol ({\small .})}
\def\bsmbl{\setplotsymbol ({\Huge .})}
\def\tarrow{\arrow <5\tdim> [.3,.6]}
\def\barrow{\arrow <8\tdim> [.3,.6]}

%These have to do with the placing of figures
\setcounter{bottomnumber}{2}
\setcounter{topnumber}{3}
\setcounter{totalnumber}{4}
\renewcommand{\bottomfraction}{1}
\renewcommand{\topfraction}{1}
\renewcommand{\textfraction}{0}

\def\draft{\renewcommand{\label}[1]{{\quad[\sf ##1]}}
\renewcommand{\ref}[1]{{[\sf ##1]}}
\renewenvironment{thebibliography}{\section*{References}}{}
\renewcommand{\cite}[1]{{\sf[##1]}}
\renewcommand{\bibitem}[1]{\par\noindent{\sf[##1]}}}
%DRAFT MACROS - COMMENT THIS OUT IN FINAL VERSION
%\draft

%Move the definition below past the style to get (c.s.e) equations
\def\theequation{\thesection.\arabic{equation}}
\preprint{\#HUTP-00/A029\\ 7/00}
\title{Chiral fermions, orbifolds, scalars and fat
branes\thanks{Research supported in
part by the
National Science Foundation
under grant number NSF-PHY/98-02709.}}
\author{
Howard~Georgi,\thanks{georgi@physics.harvard.edu}
Aaron~K.~Grant,\thanks{grant@gauss.harvard.edu}
Girma~Hailu\thanks{hailu@feynman.harvard.edu}
\\
Lyman Laboratory of Physics \\
Harvard University \\
Cambridge, MA 02138
}
\date{7/00}
\abstract{We note that orbifold boundary conditions that produce chiral
fermion zero modes in compactified higher dimensional theories may
distort scalar field vacuum expectation values, giving rise to
nontrivial dependence on the extra dimensions. We illustrate this in a
simple five dimensional model which has chiral fermion zero-modes stuck
to fat branes. The model could provide a simple and explicit realization
of the separation of quarks and leptons in the fifth dimension. We
discuss the KK expansion in some detail. We find that there are in
general non-zero-mode states stuck to the brane, like the chiral zero
modes. We see explicitly the transition from states dominated by the
internal structure of the fat brane to those dominated by the
compactification.}

\starttext

\setcounter{equation}{0}
\section{Chiral fermions in five dimensions\label{simple}}

A field theory in a space of more than four space-time dimensions may be
relevant to the description of the real world if the extra dimensions
are compactified~\cite{Arkani-Hamed:1998rs,Antoniadis:1990ew}. 
It may even be relevant if
the extra dimensions are infinite, if the gravitational interactions
distort the extra dimensions in an appropriate
way~\cite{Randall:1999vf}. We will ignore the gravitational
interactions. Our starting point will be an effective description,
approximately valid at long distance, of a theory with compactified
extra dimensions.
We will focus on the chiral orbifold boundary conditions that seem
necessary to obtain chiral fermion zero-modes from the compactified
extra dimensions~\cite{Dienes:1999vg,Cheng:1999bg}. The basic point is
simple and generic. If the orbifold boundary conditions force a scalar
field to be odd at an orbifold fixed-point, the vacuum expectation value
(VEV) must vanish on the fixed-point. If the potential is such that the
field develops a VEV in the interior, a nontrivial shape must result for
the VEV. We will describe in detail a simple model in which the orbifold
boundary condition clashes with the tendency for a scalar field to
develop a constant VEV. The result is a nontrivial
model of a fat brane that supports chiral
fermion zero-modes in a larger compactified space. We will be able to
analyze the Kaluza-Klein (KK) expansion for this system in quantitative
detail. We will find that there are in general non-zero-mode states
stuck to the brane, like the chiral zero modes. We will see explicitly
the transition from these states dominated by the internal structure of
the fat brane to those dominated by the compactification.

Our starting point is a simple example of a chiral orbifold boundary
condition equivalent to a model discussed in \cite{Cheng:1999bg}.
Consider a free massless fermion field in five dimensions in which the
extra dimension, $x_5$, is in the interval $[0,L]$. The Lagrangian is\footnote{Note that we are ignoring the possibility of interactions on the boundaries at $x_5=0$ and $L$. This is dangerous even in an effective field theory treatment because interactions in the bulk may induce interactions on the boundaries. However, we believe that our conclusions are unaffected, and we will return to the general issue in a separate paper. We are grateful to Misha Voloshin for discussions of this issue.}
\begin{equation}
\cL=\bar\psi\bigl(i\spur\,\partial-\gamma_5\partial_5\bigr)\psi\,.
\label{f1}
\end{equation}
The field $\psi$ has four components and at the Lagrangian level the
theory appears vector-like. However, we will impose boundary conditions
on the field that are periodic up to a $Z_2$ symmetry of the Lagrangian,
so that the extra dimension becomes an
orbifold.\footnote{Cumrun Vafa has emphasized to us the differences
between our construction and a string theory orbifold. He notes that
because we miss the ``winding'' modes that are stuck to the fixed-points,
our procedure may be quite dangerous, possibly leading to non-unitary
theories. We do not see how such disasters can occur in the effective
field theory approach we take in this paper. However, the reader should
be warned that our examples may be difficult to reproduce in a more
fundamental scheme such as string theory.} In the process, we will
introduce some chiral structure.
The Lagrangian (\ref{f1}) is invariant under the
transformation\footnote{Note that the masslessness of the fermion is
important --- a constant mass term would not be invariant under
(\ref{f2}). However, there is a singular limit of the model we 
discuss that corresponds to a mass term that is piecewise 
continuous with a discontinuity on the orbifold boundary.}
\begin{equation}
\psi(x,x_5)\rightarrow\Psi(x,x_5)\equiv\gamma_5\psi(x,L-x_5)\,.
\label{f2}
\end{equation}
With this $Z_2$ symmetry in hand, we can impose modified periodic
conditions on our fermion field in the following form:
\begin{equation}
\psi(x,x_5)=\Psi(x,L+x_5)=\psi(x,2L+x_5)\,.
\label{f0}
\end{equation}
That is, the field is periodic in $x_5$ with period $2L$, but if $x_5$
is translated by an odd multiple of $L$, one gets not $\psi$, but the
transformed field, $\Psi$. It is through this boundary condition that
chirality enters into the theory~\cite{Cheng:1999bg}. Specifically,
(\ref{f0}) implies the following behaviors near $x_5=0$ and $L$,
\begin{equation}
\psi(x,-x_5)=\Psi(x,L-x_5)=\gamma_5\psi(x,x_5)\,,\quad
\psi(x,L+x_5)=\Psi(x,x_5)=\gamma_5\psi(x,L-x_5)\,.
\label{f3}
\end{equation}
(\ref{f3}) shows that the points $x_5=0$ and $L$ are fixed-points of the orbifold boundary conditions.
If we decompose $\psi$ into chiral components, $\psi_\pm$, where
\begin{equation}
\psi=\psi_++\psi_-\,,\quad \gamma_5\psi_\pm=\pm\psi_\pm\,,
\label{f4}
\end{equation}
then (\ref{f3}) is equivalent to having the chiral fields defined on a
circle, $x_5\in[0,2L)$ with $2L$ identified with $0$, but with the
chiral components $\psi_\pm$ required to be respectively symmetric and
antisymmetric at the fixed-points $x_5=0$ and $x_5=L$, so this is an $S_1/Z_2$
orbifold~\cite{Pomarol:1998sd}.

Obviously, this simple model has a chiral fermion zero-mode,
\begin{equation}
\psi_+(x,x_5)=\psi(x)\,,\quad
\psi_-(x,x_5)=0\,,
\label{f5}
\end{equation}
independent of the extra dimension. All the non-zero-modes come in
chiral pairs, as they must. In this simple case, we can find them
explicitly with ease. In general, the modes of mass $M$ have the form
\begin{equation}
\psi_{M+}(x,x_5)=\psi_{M+}(x)\xi_{M+}(x_5)\,,
\quad
\psi_{M-}(x,x_5)=\psi_{M-}(x)\xi_{M-}(x_5)\,,
\label{f6}
\end{equation}
where
\begin{equation}
-\partial_5\xi_{M-}=M\xi_{M+}\,,\quad\partial_5\xi_{M+}=M\xi_{M-}\,,
\label{f7}
\end{equation}
and $\xi_{M\pm}(x_5)$ are respectively symmetric and antisymmetric at 
the
points $x_5=0$ and $x_5=L$.
For non-zero $M$, we can change the sign of $M$ by simply changing the
sign of $\xi_{M-}$. Solving (\ref{f7}) gives
\begin{equation}
\xi_{M+}(x_5)=k\cos n\pi x_5/L\,,\quad
\xi_{M-}(x_5)=-k\sin n\pi x_5/L\,,
\label{f8}
\end{equation}
where
\begin{equation}
M=n\pi/L
\label{f9}
\end{equation}
and $k$ is a normalization factor.

This simple model has a chiral zero-mode that is uniformly spread over
the compact extra dimension. In section~\ref{model}, we show that when
we add a scalar field to the model in a simple and obvious way, we
produce zero-modes that are concentrated near the orbifold fixed-point.
The reason is that we add a potential that produces a non-zero VEV for
the scalar field that breaks the symmetry between the two orbifold
fixed-points. Furthermore, the orbifold boundary conditions make it
impossible for the VEV to be constant. The generic result is a pair of
fat branes with a highly nontrivial structure in the fifth dimension
whose consequences we explore in the rest of the paper. In
section~\ref{fat}, we discuss the KK expansion for the scalar field and
a fermion field. Using techniques borrowed from supersymmetric quantum
mechanics~\cite{susy}, we construct many of the KK modes in detail and
identify qualitative features that depend on the nontrivial structure in
the extra dimension.

\setcounter{equation}{0}
\section{Scalars and their VEVs\label{model}}

The elaborated model lives in the same five dimensional space as the
previous model, and involves a single additional real scalar field,
$\phi$. The Lagrangian is
\begin{equation}
\cL=\bar\psi\bigl(i\spur\,\partial-\gamma_5\partial_5-f\phi\bigr)\psi
+\half\partial^\mu\phi\partial_\mu\phi
-\half\partial_5\phi\partial_5\phi
-{\lambda\over4}\bigl(\phi^2-v^2\bigr)^2
\label{1}
\end{equation}
where the couplings $f$ and $\lambda$ are real. The Lagrangian (\ref{1})
is invariant under the transformation
\begin{equation}
\phi(x,x_5)\rightarrow\Phi(x,x_5)\equiv-\phi(x,L-x_5)\,,
\quad
\psi(x,x_5)\rightarrow\Psi(x,x_5)\equiv\gamma_5\psi(x,L-x_5)\,,
\label{2}
\end{equation}
Now as before we can require modified periodic boundary conditions
\begin{equation}
\psi(x,-x_5)=\Psi(x,L-x_5)=\gamma_5\psi(x,x_5)\,,\quad
\psi(x,L+x_5)=\Psi(x,x_5)=\gamma_5\psi(x,L-x_5)\,,
\label{3}\\
\end{equation}
\begin{equation}
\phi(x,-x_5)=\Phi(x,L-x_5)=-\phi(x,x_5)\,,\quad
\phi(x,L+x_5)=\Phi(x,x_5)=-\phi(x,L-x_5)\,.
\label{4}
\end{equation}

The boundary conditions, (\ref{4}), require that the scalar field vanish
on the orbifold fixed-points at $x_5=0$ and $L$. However, if $v^2>0$ in
(\ref{1}), the scalar field wants to develop a vacuum expectation value.
The result is that if $\lambda v^2$ is sufficiently large there is a
minimum energy configuration in which
\begin{equation}
\left\langle\phi(x,x_5)\right\rangle=h(x_5)\,,
\label{5}
\end{equation}
where the real function $h(x_5)$ satisfies
\begin{equation}
h(0)=h(L)=0\,,\quad h(x_5)=h(L-x_5)>0\,\mbox{ for }\,0<x_5<L\,.
\label{6}
\end{equation}
There is another solution with
\begin{equation}
\left\langle\phi(x,x_5)\right\rangle=-h(x_5)\,,
\label{5e}
\end{equation}
related to (\ref{5}) by the symmetry transformation, (\ref{2}).

Now the fermion modes are given by (\ref{f6}) where the $\xi$s satisfy
\begin{equation}
\bigl(-\partial_5+f\,h(x_5)\bigr)\xi_{M-}=M\xi_{M+}\,,\quad
\bigl(\partial_5+f\,h(x_5)\bigr)\xi_{M+}=M\xi_{M-}
\label{7}
\end{equation}
with $\xi_{M\pm}(x_5)$ respectively symmetric and antisymmetric at the
points $x_5=0$ and $x_5=L$.

The non-zero-modes of (\ref{7}) come in pairs as always. For every mode
with $M=\mu\neq0$, we can always find a mode with $M=-\mu$ by changing
the sign of $\xi_{M-}$. However, there is a chiral zero-mode that must 
have
$\xi_{0-}=0$ because the boundary conditions and the differential 
equations
for $M=0$ cannot be satisfied simultaneously for non-zero $\xi_{0-}$. 
The
zero-mode looks like
\begin{equation}
\xi_{0+}(x_5)=k\,e^{-s(x_5)}\,,\quad\xi_{0-}(x_5)=0\,,
\label{8}
\end{equation}
for
\begin{equation}
s(x_5)=f\int_0^{x_5}\,dy\,h(y)\,.
\label{9}
\end{equation}
Note that the boundary conditions (\ref{3}) are automatically satisfied
because $h(x_5)$ vanishes on the fixed-points at $x_5=0$ and $x_5=L$.
If instead we tried to find a non-zero solution for $\xi_{0-}$, we would
have
to choose the normalization constant $k$ equal to zero to satisfy the
boundary conditions, so no non-trivial solution is possible.

If $f\,h(x_5)$ is positive, the zero-mode in (\ref{8}) is concentrated
at $x_5=0$. If it is negative, the zero-mode is concentrated at $x_5=L$.
If there are several fermions with couplings of different signs, those
with positive couplings will be concentrated at $x_5=0$ while those with
negative couplings will be concentrated at $x_5=L$. Thus this 
could give a very
simple explicit realization of the idea 
of~\cite{Arkani-Hamed:2000dc,Arkani-Hamed:1999za}
that if quarks and leptons are localized on different branes, the proton
can be stabilized.

In the free fermion example of section~\ref{simple}, the limit
$L\rightarrow\infty$ is singular because the zero-mode is not
normalizable in the limit. However, in the model of (\ref{1}) with
$f\,h(x_5)>0$, with the normalizable zero-mode stuck to the orbifold
fixed-point at the origin, we can take $L\rightarrow\infty$ without
doing violence to the physics. In fact, the theory simplifies in this
limit. This simple $L=\infty$ limit is not particularly interesting
phenomenologically. If we were to couple gauge fields to the fermions,
as we must certainly do to get a realistic model, taking
$L\rightarrow\infty$ would send the effective gauge coupling to the
fermion zero-modes to zero (because the gauge fields would be spread
over the whole space and the four-dimensional gauge coupling would to go
zero like $1/\sqrt{L}$). However, the $L=\infty$ theory is a very
interesting toy model, because we can do the KK expansion explicitly.
Thus we will discuss the $L=\infty$ theory to help us understand the
more interesting case of finite but large $L$.

In the $L=\infty$ theory, the orbifold is a half line which is the real
line modded out by the $Z_2$:
\begin{equation}
\phi(x,x_5)\rightarrow -\phi(x,-x_5)\,,
\quad
\psi(x,x_5)\rightarrow \gamma_5\psi(x,-x_5)\,.
\label{inf2}
\end{equation}
We will see in the next section that the KK modes in the $L=\infty$
model can be found analytically.~\cite{voloshin}

\setcounter{equation}{0}
\section{Fat branes\label{fat}}

In this section, we examine the model of section~\ref{model} in more
quantitative detail. Because the shape of the fermion zero-mode in
(\ref{8}) defines a particular ``fat brane,'' it may be interesting to
identify effects that depend on the finite extent of the
zero-mode. One such effect arises when we integrate out the scalars. We
will get 4-fermion operators with calculable coefficients. These and
other such effects depend on the structure of the KK modes. Here we
discuss the
KK expansion. We will see that we can find the form of the modes
explicitly in the limit $L\rightarrow\infty$. That in turn will allow us
to write an excellent approximation to many of the modes for large
finite $L$.

We are interested in the Lagrangian as a function of the shifted field,
\begin{equation}
\tilde\phi(x,x_5)\equiv\phi(x,x_5)-h(x_5)\,.
\label{b0}
\end{equation}
The function $h(x_5)$ is the value of $\varphi(x_5)$ that minimizes
\begin{equation}
\int_0^L\,dx_5\,\left(\half\partial_5\varphi(x_5)\partial_5\varphi(x_5)
+{\lambda\over4}\bigl(\varphi(x_5)^2-v^2\bigr)^2\right)
\label{b1e}
\end{equation}
subject to the boundary condition
\begin{equation}
\varphi(0)=\varphi(L)=0\,.
\label{b1eb}
\end{equation}
The function satisfies
\begin{equation}
\partial_5^2h(x_5)=-\lambda(v^2-h(x_5)^2)\,h(x_5)\,.
\label{b1}
\end{equation}
Evidently, there is a tension between the boundary conditions that force
the field to vanish on the orbifold fixed-points and the potential
which wants to produce a VEV in the interior.

Before we discuss the form of $h(x_5)$, let us consider the constraints
on the parameters, $L$, $f$, $\lambda$ and $v$ in the effective field
theory.
It is important to note that the various dimensional parameters in the
effective low energy theory are not {\it a priori} related. All come
down to us from some more fundamental theory at shorter distances, and
each of the effective theory parameters must satisfy a constraint in
order that the effective theory make sense. But they need not be related
to each other. This will be important to us because we will find a
region in parameter space in which the calculation is particularly
simple and transparent. The constraints from effective field theory are
simply that the dimensional parameters are small (or large) compared to
the fundamental scale to the appropriate power. Thus if the fundamental
scale is $M_P$, the length $L$ is much greater than $1/M_P$, the Yukawa
coupling $f$ is much smaller than $1/\sqrt{M_P}$, the self-coupling
$\lambda$ is much smaller than $1/M_P$, and the mass $\lambda v^2$ is
much smaller than $M_P^2$. This is summarized in equation (\ref{sum}):
\begin{equation}
L\gg{1\over M_P}
\,,\quad
f\ll{1\over\sqrt{M_P}}
\,,\quad
\lambda\ll{1\over M_P}
\,,\quad
\lambda v^2\ll M_P^2\,.
\label{sum}
\end{equation}
But for example the dimensionless quantity $\lambda v^2 L^2$ is not
constrained.

For generic values of the parameters $\lambda, v, L, \dots$, it is
difficult to study the model analytically.  However, in the limit
\begin{equation}
\label{LargeR}
L^2  \gg \frac{1}{\lambda v^2},
\end{equation}
it is relatively easy to construct approximate solutions.  In this case,
the solution for $h(x_5)$ can be approximated by a series of well
separated
kinks.

First consider $L=\infty$. Then a solution to (\ref{b1})
is a single kink given by
\begin{equation}
h(x_5) = v \tanh \sqrt{\lambda\over2} v x_5\,.
\label{k1}
\end{equation}
It is convenient to choose units in which
\begin{equation}
2\lambda\,v^2=1
\label{units}
\end{equation}
because $2\lambda\,v^2$ is the mass parameter that determines the
physical size of the kink. In these units, (\ref{LargeR}) becomes simply
$L\gg1$ and (\ref{k1}) is
\begin{equation}
h(x_5) = v \tanh {x_5\over2}\,.
\label{k1i}
\end{equation}

For large finite $L$, we can construct approximate solutions by putting
together kinks at $x_5=0$ and $x_5=L$.  On the interval $[0,L]$,
such a solution can be accurately approximated by
\begin{equation}
h(x_5) \simeq v \tanh {x_5\over2}
\tanh {L-x_5\over2}
+{\cal{O}}(e^{-L})\,.
\label{approxh}
\end{equation}
The VEV, $h(x_5)$, is odd about each of the orbifold fixed points and
can be
continued to all values of $x_5$ subject to the orbifold
boundary conditions.

	Fluctuations of $\phi(x)$ about $\langle \phi(x) \rangle = h(x_5)$
can be studied using a KK expansion.  We write
\begin{equation}
\label{ScalarKK}
\phi(x) = h(x_5) + \tilde\phi(x) = h(x_5) + \sum_{n} \phi_n(x^\mu)
f_n(x_5),
\end{equation}
where the $\phi_n(x^\mu)$ depend only on the four coordinates of the
non-compact space.  We normalize the $f_n$ to unity:
\begin{equation}
\int dx_5 \, f_n^2(x_5) = 1.
\end{equation}
Subtituting (\ref{ScalarKK}) into the action for the scalars from
(\ref{1}) and
expanding to quadratic order in $\phi_n$, we find
\begin{equation}
\label{ScalarKKModeAction}
S = \int d^4 x  \sum_n \Biggl\{ \frac{1}{2}
(\partial \phi_n)^2 -  \frac{1}{2} \biggl[ \int d x_5\,
f_n ( -f_n'' + m^2(x_5) f_n ) \biggr] \phi_n^2 \Biggr\},
\end{equation}
where
\begin{equation}
m^2(x_5) = \frac{\partial^2 V(\phi)}{\partial\phi^2}
\biggl|_{\phi = h(x_5)}\,.
\label{KKmass}
\end{equation}
For infinite $L$
\begin{equation}
m^2(x_5)= \biggl\{ 1 - {3\over2}
\sech^2{x_5\over2}\biggr\}
\label{KKRinfinite}
\end{equation}
while for large finite $L$
\begin{equation}
m^2(x_5)\simeq \biggl\{ 1 -  {3\over2}
\sech^2{x_5\over2}
-{3\over2}
\sech^2{L-x_5\over2} \biggr\}\,.
\label{KKRlarge}
\end{equation}
If the KK modes $f_n$ are chosen to satisfy the equivalent
Schr\"odinger eigenvalue problem
\begin{equation}
-f_n'' + m^2(x_5) f_n = m_n^2 f_n,
\end{equation}
then (\ref{ScalarKKModeAction}) reduces to the action for an infinite
number of four-dimensional scalar particles of mass $m_1,m_2,\dots$.

We can understand the KK spectrum by considering the case of an
infinite extra dimension.  The Schr\"odinger equation in this limit
becomes
\begin{equation}
\label{ScalarSchrodinger}
-f_n'' +\left( 1 - {3\over2}\, \sech^2 {x_5\over2} \right) f_n
= m_n^2 f_n.
\end{equation}
On the infinite line, this Schr\"odinger equation possesses two bound
states, with $m^2=0$ and
$m^2=3 / 4$, with wavefunctions
\begin{equation}
f_0(x_5)\propto\sech^2 {x_5\over2}\,,\quad
f_1(x_5)\propto\sinh {x_5\over2}
\sech^2 {x_5\over2}\,.
\label{KKmodes}
\end{equation}
As expected, the ground state is even about $x_5=0$ and the excited
state is odd.
What is going on here is that the zero-mode is associated with the
translational symmetry of the infinite case. There is a zero-mode
because in the theory on the infinite line, the kink is free to sit
anywhere. In our theory, however, translation invariance in the $x_5$
direction is broken by the boundary conditions. This kink is stuck to
the orbifold and there is no scalar zero-mode. This is consistent
because the solution $f_0$ is ruled out by our boundary condition that
$\phi$ be odd in $x_5$ at the origin. Thus $f_1$ is the only discrete KK
mode for $L=\infty$. Evidently, since it exists for $L=\infty$, it is
associated with the fat brane rather than the compactification. The
remaining solutions are continuum states
with $m^2 \geq 1$ (again in units with $2\lambda v^2=1$).  The continuum
state with $m^2=k^2+1$ has the form
\begin{equation}
f_{2k}(x_5)\propto\Biggl( \frac{1}{2} - k^2 - \frac{3}{4} \sech^2
\frac{x_5}{2}
\Biggr) \sin k x_5
- \frac{3 k}{2} \tanh\frac{x_5}{2} \cos k x_5\,.
\label{continuum}
\end{equation}
All these results are derived in detail in appendix~\ref{susy}.

Returning to the case of a finite extra
dimension, we expect the solutions to look like solutions to the
infinite problem near the orbifold fixed-points. Because the normalized
solution $f_1(x_5)$ goes to zero as $x_5\rightarrow\infty$, we get
approximate solutions $m^2 \simeq 3/4 (= 3\lambda v^2/2)$ for large $L$
by taking linear combinations of copies of this mode
centered at $x_5=0$ and $x_5=L$.
There will be two such states, corresponding to the ``plus'' and
``minus''
linear combinations of the wavefunctions centered at $x_5=0$ and
$x_5=L$,
\begin{equation}
f_1(x_5)\pm f_1(L-x_5)\,.
\label{finite1}
\end{equation}
In addition, we expect the usual KK ``continuum'' states, with
masses above $2\lambda v^2$ spaced by $\Delta m \simeq \pi / L$.
Approximate solutions for these can be obtained from (\ref{continuum})
by taking
\begin{equation}
f_{2k}(x_5)\;\;\mbox{for $x_5<L/2$}
\label{fc1}
\end{equation}
and either
\begin{equation}
f_{2k}(L-x_5)\;\;\mbox{for $x_5>L/2$ for $k$ such that $
f'_{2k}(L/2)=0$,}
\label{fc2}
\end{equation}
or
\begin{equation}
-f_{2k}(L-x_5)\;\;\mbox{for $x_5>L/2$ for $k$ such that $
f_{2k}(L/2)=0$.}
\label{fc3}
\end{equation}
The matching conditions at $x_5=L/2$ then approximately determine the
allowed $k$s.

For the fermions, the modes of mass $M$ must satisfy
\begin{equation}
a\,\xi_{M+}=M\,\xi_{M-}\,,\quad\quad
a^\dagger\,\xi_{M-}=M\,\xi_{M+}
\label{fm1}
\end{equation}
where
\begin{equation}
a=\partial_5+f\,h(x_5)\,,\quad
a^\dagger=-\partial_5+f\,h(x_5)\,.
\label{fm2}
\end{equation}
As with the scalars, we can make exact statements about these modes in
the case $L=\infty$, and reliable approximate statements for large but
finite $L$. We will simply state results here. Some details are in
appendix~\ref{fermions} and more will appear in \cite{girma}.

For $L=\infty$, (\ref{fm1}) and (\ref{fm2}) become
\begin{equation}
a_w\,\xi_{M+}=M\,\xi_{M-}\,,\quad\quad
a_w^\dagger\,\xi_{M-}=M\,\xi_{M+}
\label{a0}
\end{equation}
where
\begin{equation}
a_w=\partial_5+w\tanh{x_5\over2}\,,\quad
a^\dagger_w=-\partial_5+w\tanh{x_5\over2}
\label{a1}
\end{equation}
with
\begin{equation}
w=f\,v>0\,.
\label{fm3}
\end{equation}
The condition (\ref{fm3}) is necessary to ensure that the normalizable
fermion zero-mode is stuck to the orbifold at $x_5=0$ so that $L$ can be
taken to infinity.\footnote{Note also that (as Misha Voloshin pointed out to us) if $f$ is large, radiative corrections may be important in the calculation of the VEV of $\phi$. We ignore this issue in this paper.}

There is always a normalizable chiral zero-mode stuck to the brane given
by (\ref{8}) and (\ref{9}), which in this case becomes simply
\begin{equation}
\xi_{M+}(x_5)\propto
g_{0,w}(x_5)\equiv\sech^{2w}{x_5\over2}\,,\quad\quad\xi_{M-}(x_5)=0\,.
\label{fm4}
\end{equation}
However, we find that for $j-1<w\leq j$ for positive integer $j$, there
are $j-1$ massive modes stuck to the brane. Like the zero mode, these
states are associated with the fat brane itself and not the
compactification. They have the form
\begin{equation}
\xi_{M+}(x_5)\propto g_{\ell,w}(x_5)\,,\quad\quad
\xi_{M-}(x_5)\propto a_w\,\xi_{M+}(x_5)\,,
\label{fm5}
\end{equation}
with
\begin{equation}
M^2=2w\ell-\ell^2\,.
\label{fm5m}
\end{equation}
The function $g_{\ell,w}$ is obtained by acting with $2\ell$
$a^\dagger$s with decreasing $w$ values on $g_{0,w-\ell}$,
\begin{equation}
g_{\ell,w}=\overbrace{\overbrace{a_w^\dagger a_{w-1/2}^\dagger}\cdots
\overbrace{a_{w-\ell+1}^\dagger a_{w-\ell+1/2}^\dagger}}^{\ell\rm\;
pairs}\,g_{0,w-\ell}(x_5)
\label{fm6}
\end{equation}
for $\ell=1$ to $j-1$.

There are a few things worth noticing about these solutions.
\begin{itemize}
\item All the functions $g_{\ell,w}(x_5)$ are even for $x_5\rightarrow-
x_5$ so the boundary condition at $x_5=0$ is satisfied.
\item $g_{\ell,w}(x_5)$ goes to zero like $e^{-(w-\ell)x_5}$ as
$x_5\rightarrow\infty$. The function inherits this behavior from
$g_{0,w-\ell}$. The $a^\dagger$s acting on it do not affect the leading
exponential behavior. This is one reason why we cannot go beyond
$\ell=j-1$ in (\ref{fm6}) --- the resulting functions would grow at
infinity and would not be normalizable.
\item $\xi_{M-}(x_5)$ is also proportional to the product in (\ref{fm6})
with the initial $a_w^\dagger$ removed because this state is an
eigenstate of $a_wa_w^\dagger$.
\end{itemize}

For $M^2>w^2$, we find continuum solutions. These cannot be written in
elementary closed form except for integer or half-integer $w$. But they 
can be found in terms of hypergeometric functions~\cite{voloshin,girma}.

Returning to the finite case, one might worry that because of the
asymmetry between $x_5=0$ where the fermions are bound and $x_5=L$ where
they are repelled, it might be difficult to find modes corresponding to
the massive normalized states in (\ref{fm6}) that satisfy the boundary
conditions at large $L$. The boundary condition is automatic for the
zero-mode, but not for the massive modes. Fortunately, there is a simple
way to construct approximate eigenfunctions for large $L$.

One way to find the normalizable chiral zero-mode for finite $L$ is to
use (\ref{8}) and (\ref{9}) with our approximate form for $h(x_5)$,
(\ref{approxh}). But another way is to think of dividing the orbifold
into two regions as we did for the scalar modes, $x_5<L/2$ dominated by
the fixed-point at $x_5=0$, and $x_5>L/2$ dominated by the fixed-point
at $x_5=L$. Near $x_5=L$, the solution looks like a non-normalizable
zero-mode, just the inverse of $g_{0,w}$
\begin{equation}
\xi_{M+}(x_5)\propto \tilde g_{0,w}(x_5)\equiv\cosh^{2w}{L-
x_5\over2}\,,\quad\quad\xi_{M-}(x_5)=0\,.
\label{rfm4}
\end{equation}
Now as $x_5$ decreases from $x_5=L$ toward $x_5=L/2$, $\tilde g_{0,w}$
increases exponentially, and at $x_5=L/2$ it can be matched to an
excellent approximation onto $g_{0,w}(x_5)$ which is exponentially
falling at the same rate.

For the $j-1$ massive modes of (\ref{fm5})-(\ref{fm6}), a similar
strategy can be applied. There are non-normalizable solutions that are
analogous to the normalizable modes at the other end of the orbifold,
\begin{equation}
\xi_{M+}(x_5)\propto \tilde g_{\ell,w}(x_5)\,,\quad\quad
\xi_{M-}(x_5)\propto a_w\,\xi_{M+}(x_5)\,.
\label{rfm5}
\end{equation}
\begin{equation}
\tilde g_{\ell,w}=\overbrace{\overbrace{a_w^\dagger a_{w-
1/2}^\dagger}\cdots
\overbrace{a_{w-\ell+1}^\dagger a_{w-\ell+1/2}^\dagger}}^{\ell\rm\;
pairs}\,\tilde g_{0,w-\ell}(x_5)
\label{rfm6}
\end{equation}
for $\ell=1$ to $j-1$.
These satisfy the boundary condition at $x_5=L$ and match smoothly onto
(\ref{fm5})-(\ref{fm6}) at $x_5=L/2$. Note that the $\tilde g$s never
vanish. Thus all the nodes in these wave functions are near the
fixed-point at $x_5=0$, as expected.

We do not know a similarly simple approximation to match the continuum
modes from the two sides of the orbifold.

\setcounter{equation}{0}
\section{Zero-modes near the fixed-point\label{near}}

The fermion zero-modes described in section~\ref{model} are all concentrated on one of the orbifold fixed-points, at $x_5=0$ or $L$. In this section, we note that by elaborating the model slightly, we can produce zero-modes that are maximized near but not on the fixed-points. Consider the following Lagrangian:\footnote{This Lagrangian was suggested to us by Martin Schmaltz, to replace a more complicated scheme that we used to get the same final result.}
\begin{equation}
\cL=\bar\psi\Bigl(i\spur\,\partial-\gamma_5\partial_5-f\Bigl[1 -a(\partial_\mu\partial^\mu-\partial_5^2)\Bigr]\phi\Bigr)\psi
+\half\partial^\mu\phi\partial_\mu\phi
-\half\partial_5\phi\partial_5\phi
-{\lambda\over4}\bigl(\phi^2-v^2\bigr)^2\,.
\label{near1}
\end{equation}
We have added to (\ref{1}) only the single term proportional to the new parameter $a$. Now, however, the zero mode is given by (\ref{8},\ref{9}) with a different $h(x_5)$ that includes the effect of the new term. For large $L$ we can write
\begin{equation}
h(x_5)\approx fv\,\left(\tanh{x_5\over2}
+a\,\partial_5^2\tanh{x_5\over2}
\right)
\,\left(\tanh{L-x_5\over2}
+a\,\partial_5^2\tanh{L-x_5\over2}
\right)
\label{near6}
\end{equation}
for constants $f$ and $a$.
This is interesting because for $a>2$, $h(x_5)$ changes sign at
\begin{equation}
\tilde x_5=2\arctanh\sqrt{1-2/a}
\label{near7}
\end{equation}
and at $x_5=L-\tilde x_5$.
For $f>0$, this describes a fermion concentrated at $\tilde x_5$ or $L-\tilde x_5$, depending on the sign of $fv$.

Note that for reasonable values of $a$, $\tilde x_5$ is of order one. Thus the zero mode does not stray very far from the orbifold fixed point. What we would expect in a model with finite $L$ and several fermions is that the fermions would fall into four sets:
{\renewcommand{\arraystretch}{1}
\begin{equation}
\begin{array}{l}
\mbox{for $f>0$ and $a<2$ the zero-mode is concentrated at $x_5=0$;}\\
\mbox{for $f>0$ and $a>2$ the zero-mode is concentrated at $x_5=\tilde x_5$ near $x_5=0$;}\\
\mbox{for $f<0$ and $a<2$ the zero-mode is concentrated at $x_5=L$;}\\
\mbox{for $f<0$ and $a>2$ the zero-mode is concentrated at $x_5=L-\tilde x_5$ near $x_5=L$.}\\
\end{array}
\label{near8}
\end{equation}
}

One annoying thing about this is that the higher derivative coupling w have added is higher dimension than the ordinary Yukawa coupling, and therefore we might expect the parameter $a$ to be small --- of order $\lambda v^2/M_P^2$. We can consistently take $a$ of order 1 in our units only if $f$ is small.
This could be a problem in model building.

\setcounter{equation}{0}
\section{Concluding questions\label{conclusions}}

We have shown in a very explicit example how scalar VEVs and orbifold
boundary conditions combine to produce nontrivial structure in the extra
dimensions. This behavior is generic, and we expect behavior of this
kind to appear in other explicit models of fat branes. The most
interesting general result is that the KK expansion may produce two
kinds of massive modes --- those truly associated with the
compactification and those stuck to the fat brane. Let us close with a
couple of very different questions.

Question: Do the KK states that are normalizable in the $L=\infty$ limit
play a special role, or are they simply the lightest of the KK
excitations? It seems likely to us that the answer is the former. These
states are stuck to the fat brane and are thus very different from the
KK states associated with compactification. Particularly intriguing is
the nearly degenerate pair of scalars in (\ref{finite1}). These may be
an important source of communication between fermions localized on the
two different branes.

Question: Does this kind of construction (which in some ways resembles
the Kaplan idea~\cite{Kaplan:1992bt}) help in any way with the
difficulties of putting chiral fermions on the lattice?  Answer: We
don't think so. It seems that this structure makes it impossible to
decouple the doublers associated with the zero-mode, but the question is
interesting and may be worth pursuing further.

\section*{Acknowledgements}

We would like to thank Nima Arkani-Hamed, Bogdan Dobrescu, Lisa
Randall,
Martin Schmaltz, John Terning, Misha Voloshin and Koichi Yamawaki
for helpful comments. HG is grateful to the Aspen Center for Physics for providing a splendid work environment while this paper was being revised. Our research is supported in part by the
National Science Foundation
under grant number NSF-PHY/98-02709.

\appendix

\setcounter{equation}{0}
\section{SUSY quantum mechanics\label{susy}}

Here we give more details about the solution for the scalar modes in the
$L=\infty$ model. We use techniques from SUSY quantum
mechanics~\cite{susy}.
In solving for the scalar -KK modes in the kink background, one
needs the eigenfunctions of the Schr\"odinger equation
(\ref{ScalarSchrodinger}).
In the case of an infinite
extra dimension, we can solve this equation exactly using supersymmetric
quantum mechanics.  First we define two sets of ``raising'' and
``lowering'' operators:
\begin{equation}
\begin{array}{cc}
a_i^\dagger = -\partial_5 + g_i(x_5), & a_i = \partial_5 + g_i(x_5)\,,
\end{array}
\end{equation}
where
\begin{equation}
g_1 = \frac{1}{2} \tanh\frac{x_5}{2}\,,\quad\quad g_2 =
\tanh\frac{x_5}{2}\,.
\end{equation}
Because
\begin{equation}
\partial_5\tanh {x_5\over2}={1\over2}\left(1-
\tanh^2{x_5\over2}\right)\,,
\end{equation}
it is straightforward to verify that
\begin{equation}
\label{H1}
a_1 a_1^\dagger
= -\partial_5^2 + {1\over4}\left(1-\tanh^2{x_5\over2}\right)
+ {1\over4}\tanh^2{x_5\over2}
= -\partial_5^2 + \frac{1}{4}\,,
\end{equation}
\begin{equation}
\label{H12}
a_1^\dagger a_1= -\partial_5^2 - {1\over4}\left(1-
\tanh^2{x_5\over2}\right)
+ {1\over4}\tanh^2{x_5\over2}
= -\partial_5^2 - \frac{1}{4}+ {1\over2}\tanh^2{x_5\over2}
\,,
\end{equation}
\begin{equation}
\label{2H1}
a_2 a_2^\dagger
= -\partial_5^2 + {1\over2}\left(1-\tanh^2{x_5\over2}\right)
+ \tanh^2{x_5\over2}
= -\partial_5^2 + {1\over2}+{1\over2}\tanh^2{x_5\over2}
\,,
\end{equation}
\begin{equation}
\label{2H12}
a_2^\dagger a_2
= -\partial_5^2 - {1\over2}\left(1-\tanh^2{x_5\over2}\right)
+ \tanh^2{x_5\over2}
= -\partial_5^2 - {1\over2}+{3\over2}\tanh^2{x_5\over2}
\,.
\end{equation}
Thus
\begin{equation}
\label{H2}
a_1^\dagger a_1 = a_2 a_2^\dagger - \frac{3}{4}
= -\partial_5^2 + \Biggl( \frac{1}{4} - \frac{1}{2}\, \sech^2
\frac{x_5}{2}
\Biggr)\,,
\end{equation}
while
\begin{equation}
\label{H3}
a_2^\dagger a_2 =
-\partial_5^2 + \Biggl( 1 - \frac{3}{2}\, \sech^2 \frac{x_5}{2}
\Biggr)\,.
\end{equation}
We see that $a_2^\dagger a_2$ is the Hamiltonian we wish to diagonalize.
To construct solutions to this Hamiltonian, we first observe that it is
trivial to find the eigenfunctions of (\ref{H1}):  these are just plane
waves.   Furthermore, given a plane wave with wave number $k$ obeying
\begin{equation}
a_1 a_1^\dagger \chi_{2k} = (k^2+1/4) \chi_{2k}\,,
\end{equation}
we can construct an eigenfunction of (\ref{H2}) by applying
$a_1^\dagger$
to both sides:
\begin{equation}
a_1^\dagger a_1 ( a_1^\dagger \chi_{2k} ) = (k^2+1/4) ( a_1^\dagger
\chi_{2k})\,.
\end{equation}
So we conclude that $a_1^\dagger a_1$ has all of the plane wave
eigenstates of
$a_1 a_1^\dagger$, plus an additional zero-energy bound state which is
obtained by solving $a_1 \chi_1 = 0$:
\begin{equation}
a_1 \chi_1 = 0 \rightarrow \chi_1 \propto \sech \frac{x_5}{2}\,.
\end{equation}
Furthermore, since (\ref{H2}) can also be expressed in terms of $a_2
a_2^\dagger$, we can use the solutions of (\ref{H2}) to find solutions
of
the final Hamiltonian (\ref{H3}).  Indeed, for each eigenvalue $\kappa$
of
$a_1^\dagger a_1$, we have
\begin{equation}
a_2 a_2^\dagger f = ( a_1^\dagger a_1 + 3/4) f
= ( \kappa^2 + 3/4) f\,.
\end{equation}
Thus
\begin{equation}
a_2 a_2^\dagger\chi_1={3\over4}\,\chi_1\,,\quad\quad
a_2 a_2^\dagger a_1^\dagger\chi_{2k}=(k^2+1)a_1^\dagger\chi_{2k}\,.
\label{a2sum}
\end{equation}
Multiplying both sides by $a_2^\dagger$ yields eigenfunctions of our
original Hamiltonian.  So the spectrum of $a_2^\dagger a_2$ consists of
all
eigenvalues of $a_1^\dagger a_1$ (shifted by $3/4$), plus a zero energy
bound state obtained from $a_2 f_0 = 0$:
\begin{equation}
a_2 f_0 = 0 \rightarrow f_0 \propto \sech^2 \frac{x_5}{2}\,.
\end{equation}
However, this zero-mode is even, and therefore does not satisfy the
boundary conditions. Thus the allowed eigenstates are $f_1$ and the odd
plane waves,
\begin{equation}
f_1 \propto a_2^\dagger \chi_1\propto a_2^\dagger\, \sech\frac{x_5}{2} =
\frac{3}{2} \tanh\frac{x_5}{2}\,\sech\frac{x_5}{2}\,,  \;\;\mbox{with
$m^2=3/4$,}
\end{equation}
and
\begin{equation}
\begin{array}{c}\displaystyle f_{2k}\propto a_2^\dagger
a_1^\dagger\chi_{2k}
\propto \Biggl( \frac{1}{2} - k^2 - \frac{3}{4} \sech^2 \frac{x_5}{2}
\Biggr) \sin k x_5
-\frac{3 k}{2} \tanh\frac{x_5}{2} \cos k x_5\,,\\
\mbox{with $m^2=k^2+1$.}
\end{array}
\end{equation}

\setcounter{equation}{0}
\section{Fermions on fat branes\label{fermions}}

Here we will sketch the proof of (\ref{fm5})-(\ref{fm6}).
From (\ref{a0})-(\ref{a1}) it follows that $\xi-$ is an eigenfunction of
\begin{equation}
a_w a_w^\dagger
= -\partial_5^2 + {w\over2}\left(1-\tanh^2{x_5\over2}\right)
+ w^2\tanh^2{x_5\over2}
\label{a2}
\end{equation}
and $\xi_{M+}$ is an eigenfunction of
\begin{equation}
a^\dagger_w a_w
= -\partial_5^2 - {w\over2}\left(1-\tanh^2{x_5\over2}\right)
+ w^2\tanh^2{x_5\over2}\,.
\label{a3}
\end{equation}
We will use a Dirac notation for the eigenfunctions, denoting an
eigenstate
of (\ref{a3}) with eigenvalue $E$ by $\Ket{E,w}$.

Now it is obvious that
\begin{equation}
a_{w-\alpha}^\dagger a_{w-\alpha}
= -\partial_5^2 - {w-\alpha\over2}\left(1-\tanh^2{x_5\over2}\right)
+ (w-\alpha)^2\tanh^2{x_5\over2}
\label{a4}
\end{equation}
and thus
\begin{equation}
a _w a_w^\dagger
-a_{w-\alpha}^\dagger a_{w-\alpha}
= +{2w-\alpha\over2}\left(1-\tanh^2{x_5\over2}\right)
+ (2w\alpha-\alpha^2)\tanh^2{x_5\over2}\,.
\label{a5}
\end{equation}
For $\alpha=1/2$, the terms proportional to $\tanh^2$ cancel in
(\ref{a5}) and we have
\begin{equation}
a_w a_w^\dagger =a_{w-1/2}^\dagger a_{w-1/2} +w -{1\over4}\,.
\label{a6}
\end{equation}

Thus if $\Ket{E-w+1/4,w-1/2}$ is an eigenstate of $a^\dagger_{w-
1/2}a_{w-1/2}$ with eigenvalue $E-w+1/4$,
then $a^\dagger_{w}\Ket{E-w+1/4,w-1/2}$ is an eigenstate of
$a^\dagger_{w}a_{w}$ with eigenvalue $E$. That is, so long as the
eigenstate $\Ket{E-w+1/4,w-1/2}$ exists, we can write
\begin{equation}
\Ket{E,w}\propto a_w^\dagger\Ket{E-w+1/4,w-1/2}\,.
\label{a7}
\end{equation}

Applying the same argument again shows that if
the eigenstate $\Ket{E-2w+1,w-1}$ exists, we can write
\begin{equation}
\Ket{E-w+1/4,w-1/2}\propto a_{w-1/2}^\dagger\Ket{E-2w+1,w-1}\,,
\label{a8}
\end{equation}
and therefore
\begin{equation}
\Ket{E,w}\propto a_w^\dagger\Ket{E-w+1/4,w-1/2}
\propto a_w^\dagger a_{w-1/2}^\dagger\Ket{E-2w+1,w-1}\,.
\label{a9}
\end{equation}
If $\Ket{E,w}$ is to be an eigenstate, this process must terminate in a
chiral zero-mode. Conversely, we get all the normalizable modes by
acting on the
zero-modes by pairs of $a^\dagger$s as in (\ref{a9}). This is the basis
of (\ref{fm6}).

\setcounter{equation}{0}
\section{Another simple limit\label{dumblimit}}

If $L$ is not large, the approximations discussed in section~\ref{fat}
fail badly. In general, we then have to resort to numerical techniques.
Here we discuss one way of approaching the problem, and identify another
simple limit.
In order to incorporate the effects of the boundary conditions, we could
expand the $\tilde\phi$ field in a set of basis functions in $x_5$. In
this case the obvious ones are
\begin{equation}
\xi_n(x_5)\equiv \sqrt{2\over L}\sin {n\pi x_5\over L}\,.
\label{b3}
\end{equation}
This will allow us to formulate the problem in general, and also make it
easy to solve it exactly in a particular limit.

Now we can formulate the problem in general by expanding $h(x_5)$ and
$\tilde\phi(x,x_5)$ in terms of the basis functions (\ref{b3}). We can
truncate this expansion for some large $n$ and solve the finite problem.
The coefficients in the expansion of $h(x_5)$ can be determined by
minimizing (\ref{b1e}). However, a close look suggests that there is a
limit of the theory in which the calculation is much simpler. One can
immediately see that the expectation value $h(x_5)$ goes to zero as
\begin{equation}
\lambda v^2 L^2\rightarrow\pi^2
\label{b4}
\end{equation}
from above. For smaller values of $\lambda v^2 L^2$ there is no vacuum
expectation value. This suggests taking
\begin{equation}
\lambda v^2 L^2=\pi^2+\epsilon
\label{b5}
\end{equation}
for small $\epsilon$. When we do that, we find that we can calculate the
coefficients in $h(x_5)$ as a power series in $\epsilon$. The first
terms are
\begin{equation}
h(x_5)=\sqrt{2\over3\lambda L}\,\epsilon^{1/2}\,\xi_1(x_5)
+{1\over24\pi^2}\sqrt{2\over3\lambda L}\,\epsilon^{3/2}\,\xi_3(x_5)
+\cO(\epsilon^{5/2})\,.
\label{b6}
\end{equation}

Furthermore, doing the KK expansion, we find that the mass squared of
the $\tilde\phi_1$ mode is $2\epsilon/L^2$, while all the other mass
squares scale with $\pi^2/L^2$, not suppressed by $\epsilon$. To leading
order, the mass squared of the mode $\tilde\phi_n$ for $n>1$ is $(n^2-
1)\pi^2/L^2$.

\end{document}